\documentclass[a4paper, oneside, twocolumn, notitlepage, 10pt]{extarticle_ecoc2015}
\usepackage{ecoc2015}

\newcommand{\GL}{\textcolor{black}}
\newcommand{\AGiA}{\textcolor{black}}
\newcommand{\AS}{\textcolor{black}}

\renewcommand{\mathbf}{\bm}

\newcommand{\Bmat}{\mathbf{B}}
\renewcommand{\b}{\boldsymbol{b}}

\renewcommand{\b}{\boldsymbol{b}}
\renewcommand{\u}{\boldsymbol{u}}

\newcommand{\ua}{\boldsymbol{u}^{\mathsf{a}}}
\newcommand{\us}{\boldsymbol{u}^{\mathsf{s}}}

\renewcommand{\P}{\mathsf{P}}
\newcommand{\mI}{\mathsf{I}}

\newcommand{\gmiHDD}{\mI_{\mathsf{HDD}}^{\mathsf{gmi}}}

\newcommand{\Rs}{R_{\mathsf{s}}}
\newcommand{\nc}{n_{\mathsf{c}}}

\newcommand{\kc}{k_{\mathsf{c}}}

\newcommand{\rmT}{\mathrm{T}}

\makeatletter
\protected\def\vvv#1{\leavevmode\bgroup\vbox\bgroup\xvvv#1\relax}

\def\xvvv{\afterassignment\xxvvv\let\tmp= }

\def\xxvvv{%
	\ifx\tmp\@sptoken\egroup\ \vbox\bgroup\let\next\xvvv
	\else\ifx\tmp\relax\egroup\egroup\let\next\relax
	\else
	\hbox to 1.1em{\hfill\tmp\hfill}
	\let\next\xvvv\fi\fi
	\next}

\makeatother

\begin{document}
	
\selectlanguage{american}    


\title{Probabilistically-Shaped Coded Modulation with\\Hard Decision Decoding for Coherent Optical Systems}


\author{
	Alireza~Sheikh\textsuperscript{(1)}, Alexandre~Graell i Amat\textsuperscript{(1)}, and
	Gianluigi~Liva\textsuperscript{(2)}
}

\maketitle                  


\begin{strip}
	\begin{author_descr}
		\vspace{-0.6cm}
		\begin{center}
			\textsuperscript{(1)} Department of Electrical Engineering, Chalmers University of Technology, \uline{asheikh@chalmers.se}\\
			\textsuperscript{(2)} Institute of Communications and
			Navigation of the German Aerospace Center (DLR)
		\end{center}
		
	\end{author_descr}
\end{strip}

\setstretch{1.1}


\begin{strip}
  \begin{ecoc_abstract}
\AGiA{We consider probabilistic shaping to maximize the achievable information rate of coded modulation (CM) with hard decision decoding.} The proposed scheme using binary staircase codes \GL{outperforms its uniform CM counterpart by more than 1.3 dB} \AGiA{for $64$-QAM and $5$ bits/symbol}.
\end{ecoc_abstract}
\end{strip}
\section{Introduction}
\GL{Coded modulation (CM) schemes allow achieving high spectral efficiencies as required by the increasing demand for network capacity. By suitably choosing the probability of occurence of the constellation points, probabilistic shaping can significantly improve the efficiency of CM. It has been recently demonstrated that a probabilistically-shaped CM scheme using DVB-S2 codes with soft decision decoding (SDD) can operate within less than $1.1$ dB of the additive white Gaussian noise (AWGN) capacity\cite{georg_tcom}.}
 
Hard decision decoding (HDD) can significantly reduce the complexity of the decoder compared to SDD. In particular, staircase codes\cite{staircase_frank}, braided codes\cite{jian2013iterative}, and generalized product codes \cite{hager_coded_mod} with HDD have been proposed for high-speed fiber-optic communications, \AGiA{yielding very large net coding gains, yet with low decoding complexity}. In this paper, we propose probabilistically-shaped CM with HDD. We consider \AGiA{a CM scheme using} binary staircase codes with Bose-Chaudhuri-Hocquenghem (BCH) as component codes \AGiA{in combination with QAM modulation}. We find the optimal distribution for the QAM constellation such that the bit-wise achievable information rate (AIR) of the \AGiA{CM system with HDD} is maximized. We also design the parameters of the staircase code such that the requested information rate is achievable. By \AGiA{means of simulations, we show that the proposed CM scheme achieves} significantly better performance compared to the \AGiA{baseline CM scheme with uniform distribution}.


\section{Preliminaries}
We consider a discrete-time additive white Gaussian noise (AWGN) channel as an approximation of the fiber-optic channel (FOC).
 We also assume a $2^m$-QAM constellation with average transmitted power $\mathbb{E}[|X_i|^{2}]\le \P$, where $|\cdot|$ denotes the absolute value. Furthermore, a block-wise transmission system is considered, \AS{where $\u$ denotes the transmitted information block, and $\hat \u$ denotes the decoded information block}. The block error probability of the system is defined as \AS{$\text{P}_\text{e} = \Pr(\u\ne\hat{\u})$}. We consider a CM scheme with binary codes and HDD. \AGiA{Let $X$ and $\hat Y$ be the random variables (RVs) corresponding to the channel input and hard detected output of the channel, respectively}. Also, let $B_X$ and $B_{\hat Y}$ be the RVs corresponding to the binary labels associated with $X$ and $\hat Y$, respectively, using the binary reflected Gray code (BRGC) labeling. An AIR for the HDD system \AGiA{is given by the generalized mutual information (GMI), which can be computed as 
 \begin{align*}
 \tiny{\gmiHDD = \mathop {\sup }\limits_{s > 0} {\mathbb{E}_{{B_X},{B_{\hat Y}}}}\left[ {{{\log }_2}\frac{{q{{\left( {{B_X},{B_{\hat Y}}} \right)}^s}}}{{{\mathbb{E}_{{B_{X^{\prime}}}}}\left[ {q{{\left( {{B_{X^{\prime}}},{B_{\hat Y}}} \right)}^s}} \right]}}} \right]},
\end{align*}}%
where $s$ is the optimization parameter and $q\left( {{B_X},{B_{\hat Y}}} \right)$ is the HDD  metric computed based on the bit-wise Hamming metric, given as $q\left( {{B_X},{B_{\hat Y}}} \right) = {\varepsilon ^{{\mathsf{d}_\mathsf{H}}\left( {{B_X},{B_{\hat Y}}} \right)}}{\left( {1 - \varepsilon } \right)^{m - {\mathsf{d}_\mathsf{H}}\left( {{B_X},{B_{\hat Y}}} \right)}}$,
 with $0 < \varepsilon < \frac{1}{2}$ and ${{\mathsf{d}_\mathsf{H}}\left( {{B_X},{B_{\hat Y}}} \right)}$ being the Hamming distance between binary labels ${B_X}$ and ${B_{\hat Y}}$.


\section{Probabilistic Shaping for the HDD System}
We consider the Maxwell-Boltzmann distribution for the channel input $X$, which maximizes the bit rate of a discrete constellation for a given average energy\cite{kschischang1993optimal}. Thus, the probability of selecting the constellation point $x \in \mathcal{X}$ is given by ${p^{\lambda}_{X}}\left( x \right) = {A_\lambda }{e^{ - \lambda {|x|^2}}}$, where $A_\lambda $ is a normalization factor and $\mathcal{X}$ is the QAM constellation alphabet.
We assume that the QAM constellation is scaled with $\Lambda$ to maintain the power constraint $\P$. Therefore, the received symbol at the output of the channel is given by ${Y} = {\Lambda X} + {Z}$, with power constraint $\mathbb{E}[(\Lambda |X|)^{2}]=\P$,
where ${Z}$ is an independent and identically distributed (i.i.d.) complex Gaussian RV with zero mean and unit variance. We find the shaping parameter $\lambda$ and scaling $\Lambda$ to maximize $\gmiHDD$, i.e., $(\Lambda^{*},\lambda^{*})=\mathop {{\text{argmax}}}\limits_{\Lambda,\lambda} \;  \gmiHDD.$
To find $(\Lambda^{*},\lambda^{*})$ for a given SNR, one can vary $\lambda$ and find the corresponding $\Lambda$ which satisfies the power constraint. Then, among all feasible pairs $(\Lambda,\lambda)$, the one which maximizes $\gmiHDD$ is selected. We denote by $p_X^{{\lambda ^*}}$ the optimized distribution. In Fig.~\ref{AIRfig}, we plot the AIR $\gmiHDD$ for uniform and shaped distribution for $64$-QAM constellation. As can be seen, the shaping significantly improves $\gmiHDD$. 
Let $\mathcal{X}^{+}$ be the set containing the $2^m$-QAM constellation points with positive real and imaginary parts (see Fig.~\ref{fig8QAM}). We denote by $A$ the RV which takes values on $\mathcal{X}^{+}$. Also, let $S_R$ and $S_I$ be the RVs correspodning to the sign of the real and imaginary part of the constellation points, respectively. It can be verified that $p_X^{{\lambda ^*}}$ is symmetric with respect to the real and imaginary axis and that the RV $X$ is represented uniquely by $A$, $S_{R}$, and $S_{I}$, i.e., $X=AS_{R}S_{I}$, with $p^{\lambda^{*}}_{A}\left( a \right) = 4p_X^{{\lambda ^*}}\left( {a} \right)$ and $p_{S_{R}}\left( 1 \right)=p_{S_{I}}\left( 1 \right) = p_{S_{R}}\left( { - 1} \right)=p_{S_{I}}\left( { - 1} \right)=\frac{1}{2}$. In Fig.~\ref{fig8QAM}, the constellation points with the same distribution are shown with the same color. Using the BRGC labeling, one can represent the points in the set $\mathcal{X}^{+}$ with $\text{log}_2(2^m/4)=m-2$ bits, since the first $2$ bits of the \AGiA{labeling of the points $x\in\mathcal{X}^{+}$ are the same. Using the mapping $0 \mapsto -1$ and $1 \mapsto +1$, the first and second bit of the labeling determine the sign of the real and  the imaginary part of the constellation points, respectively.}

\begin{figure}[t] \centering 
	\newlength\figureheight
	\newlength\figurewidth
	\setlength\figureheight{2.6cm}
	\setlength\figurewidth{8.8cm}  
%
%
\definecolor{mycolor1}{rgb}{0.49412,0.18431,0.55686}%
\definecolor{mycolor2}{rgb}{0.00000,0.49804,0.00000}%
\definecolor{mycolor3}{rgb}{0.00000,0.44706,0.74118}%
\definecolor{mycolor4}{rgb}{0.49412,0.18431,0.55686}%
\begin{tikzpicture}[scale=.8]
\pgfplotsset{every tick label/.append style={font=\Large}}
\begin{axis}[%
width=0.95\figurewidth,
height=1.47\figureheight,
at={(0\figurewidth,0\figureheight)},
scale only axis,
xmin=8,
xmax=24,
ymin=2,
ymax=6,
ylabel={\large AIR},
xlabel={\large {$\text{E}_\text{s}\text{/N}_\text{0}$ (dB)}},
ylabel style={yshift=-0.55cm},
xlabel style={yshift=0.1cm},
xmajorgrids,
ymajorgrids,
axis background/.style={fill=white},
legend style={at={(0.98,0.03)},anchor=south east,row sep=0pt, legend cell align=left,align=left,draw=black, fill=white, inner sep=1.3pt, outer sep=0.9pt}
]
\addplot [color=mycolor4,solid,line width=1.0pt]
table[row sep=crcr]{%
	5	0.999062623942368\\
	5.25	1.04161939853981\\
	5.5	1.08518319110512\\
	5.75	1.12982015705296\\
	6	1.17561171097448\\
	6.25	1.22251454568213\\
	6.5	1.27055866902754\\
	6.75	1.31977116414262\\
	7	1.37017527873563\\
	7.25	1.42196573422968\\
	7.5	1.474998580681\\
	7.75	1.52927589274995\\
	8	1.5848900270839\\
	8.25	1.64190047371988\\
	8.5	1.70016662725648\\
	8.75	1.75984076197366\\
	9	1.82087158532825\\
	9.25	1.8832208290044\\
	9.5	1.94702539426181\\
	9.75	2.01213725898699\\
	10	2.07872181166046\\
	10.25	2.14668296304102\\
	10.5	2.21605960502699\\
	10.75	2.2869616229617\\
	11	2.35928780160918\\
	11.25	2.43308955979793\\
	11.5	2.5085178111805\\
	11.75	2.58548750223494\\
	12	2.66403759299672\\
	12.25	2.7442020608791\\
	12.5	2.82600565680176\\
	12.75	2.90946161261522\\
	13	2.99456925754282\\
	13.25	3.08131162659362\\
	13.5	3.169762487932\\
	13.75	3.25978698794393\\
	14	3.35134882577708\\
	14.25	3.44442914743305\\
	14.5	3.53890215696059\\
	14.75	3.63464264476682\\
	15	3.73154062027921\\
	15.25	3.82944839339683\\
	15.5	3.92818129231541\\
	15.75	4.02751907716647\\
	16	4.12720803964009\\
	16.25	4.22705660593868\\
	16.5	4.32673101064465\\
	16.75	4.42596985615945\\
	17	4.52447405460975\\
	17.25	4.62191301301898\\
	17.5	4.7179314431442\\
	17.75	4.81220225806347\\
	18	4.90436921942489\\
	18.25	4.99409855943148\\
	18.5	5.08104338335846\\
	18.75	5.16486778164169\\
	19	5.24526587284423\\
	19.25	5.32195520307841\\
	19.5	5.39466520230525\\
	19.75	5.46316166682473\\
	20	5.52726242116612\\
	20.25	5.58681822952208\\
	20.5	5.64173697706689\\
	20.75	5.69196673426747\\
	21	5.73752042354205\\
	21.25	5.77845380319505\\
	21.5	5.81488182575306\\
	21.75	5.84696918000668\\
	22	5.87492409186845\\
	22.25	5.89899913389514\\
	22.5	5.9194774580799\\
	22.75	5.93667048961884\\
	23	5.95090514777848\\
	23.25	5.96251796794189\\
	23.5	5.97184450198895\\
	23.75	5.97921155913974\\
	24	5.98492876943471\\
	24.25	5.98928331387043\\
	24.5	5.99253473417472\\
	24.75	5.99491183370014\\
	25	5.99661129991762\\
};
\addlegendentry{$\gmiHDD$ uniform};

\addplot [color=mycolor2,solid,line width=1.0pt]
table[row sep=crcr]{%
	5	1.46579892317654\\
	5.25	1.51178911680456\\
	5.5	1.55619123984638\\
	5.75	1.59882968188983\\
	6	1.63953427325279\\
	6.25	1.68000853670076\\
	6.5	1.72577949084277\\
	6.75	1.77689532479202\\
	7	1.83272060466621\\
	7.25	1.89255005724396\\
	7.5	1.95571822430114\\
	7.75	2.02166184994724\\
	8	2.08990018738181\\
	8.25	2.1600553890047\\
	8.5	2.2318070653719\\
	8.75	2.304916488661\\
	9	2.37916205387676\\
	9.25	2.45438436936732\\
	9.5	2.5304187433945\\
	9.75	2.60715269627337\\
	10	2.68446550131693\\
	10.25	2.76226187311712\\
	10.5	2.84045015770386\\
	10.75	2.91896869850708\\
	11	2.99773143907823\\
	11.25	3.07671422861147\\
	11.5	3.15586817838582\\
	11.75	3.23517892321605\\
	12	3.31467672592584\\
	12.25	3.39435455116865\\
	12.5	3.47423829795137\\
	12.75	3.5543838786503\\
	13	3.63481492304937\\
	13.25	3.71555889895922\\
	13.5	3.79661804137871\\
	13.75	3.87795188105578\\
	14	3.95955103232285\\
	14.25	4.04130553418989\\
	14.5	4.12310827933874\\
	14.75	4.20485845514854\\
	15	4.28642427445235\\
	15.25	4.36767479289973\\
	15.5	4.44840442014369\\
	15.75	4.52852864511378\\
	16	4.60782481644679\\
	16.25	4.68615062673721\\
	16.5	4.76333932398564\\
	16.75	4.83919931168793\\
	17	4.91364046986271\\
	17.25	4.98639190833142\\
	17.5	5.05742527251708\\
	17.75	5.12647157663465\\
	18	5.19340459798565\\
	18.25	5.2581069764854\\
	18.5	5.3204322365192\\
	18.75	5.38023999842817\\
	19	5.43737553927009\\
	19.25	5.49181532197591\\
	19.5	5.54345964303734\\
	19.75	5.59206986965365\\
	20	5.63785184467143\\
	20.25	5.68050461548196\\
	20.5	5.72005698814792\\
	20.75	5.75666720425817\\
	21	5.7901578693714\\
	21.25	5.82058855324342\\
	21.5	5.84803707424999\\
	21.75	5.87258686484818\\
	22	5.8943312455479\\
	22.25	5.91337116309528\\
	22.5	5.92981285691288\\
	22.75	5.94397668986422\\
	23	5.95595535683365\\
	23.25	5.96576921435593\\
	23.5	5.97402262310257\\
	23.75	5.98049623878471\\
	24	5.98571329436761\\
	24.25	5.98975673418029\\
	24.5	5.99275855496915\\
	24.75	5.99493988371216\\
	25	5.99661129991762\\
	25.25	5.99779782206388\\
	25.5	5.99860562755232\\
	25.75	5.99914113721467\\
	26	5.9994862299573\\
};
\addlegendentry{$\gmiHDD$ shaping};

\addplot [color=mycolor3,solid,line width=1.0pt]
table[row sep=crcr]{%
	5	1.00000552003331\\
	5.25	1.00000552003331\\
	5.5	1.00000552003331\\
	5.75	1.00000552003331\\
	6	1.00000552003331\\
	6.25	1.157876928828\\
	6.5	1.32353903271756\\
	6.75	1.45675846151053\\
	7	1.56698707891994\\
	7.25	1.66481034157059\\
	7.5	1.75127224944022\\
	7.75	1.82879194997537\\
	8	1.90799255339484\\
	8.25	1.97786634205178\\
	8.5	2.05221988838854\\
	8.75	2.12555116186289\\
	9	2.19718212020344\\
	9.25	2.27290288918891\\
	9.5	2.34621743349079\\
	9.75	2.41640175686331\\
	10	2.49026900065954\\
	10.25	2.56019909195096\\
	10.5	2.64212753683267\\
	10.75	2.71120244078796\\
	11	2.78374315485465\\
	11.25	2.86009805572241\\
	11.5	2.93034976079943\\
	11.75	3.01497584354267\\
	12	3.08177279868982\\
	12.25	3.16365545178417\\
	12.5	3.23752270507571\\
	12.75	3.31507739056627\\
	13	3.39660318099579\\
	13.25	3.4823836206079\\
	13.5	3.55731160259184\\
	13.75	3.63551648123379\\
	14	3.71710210125419\\
	14.25	3.80212314129943\\
	14.5	3.8726051718061\\
	14.75	3.94522370640973\\
	15	4.01987364021476\\
	15.25	4.09638517013066\\
	15.5	4.17450940034261\\
	15.75	4.23395817465942\\
	16	4.29393770241768\\
	16.25	4.35422343656377\\
	16.5	4.41454777467145\\
	16.75	4.47459652308648\\
	17	4.51429838235083\\
	17.25	4.57305387404306\\
	17.5	4.61149787140198\\
	17.75	4.6491912047721\\
	18	4.68597685428518\\
	18.25	4.72168776583955\\
	18.5	4.75614733728094\\
	18.75	4.78917010335418\\
	19	4.82056263452022\\
	19.25	4.8355852095352\\
	19.5	4.86415511105399\\
	19.75	4.87765043371992\\
	20	4.90293031311388\\
	20.25	4.91466191331352\\
	20.5	4.93617593100207\\
	20.75	4.94590559177726\\
	21	4.95491548896304\\
	21.25	4.96317981847353\\
	21.5	4.97067314707936\\
	21.75	4.97737049407994\\
	22	4.983247414588\\
	22.25	4.98828008404869\\
	22.5	4.99244538358718\\
	22.75	4.99244538358718\\
	23	4.99572098575262\\
	23.25	4.99572098575262\\
	23.5	4.9980854402029\\
	23.75	4.9980854402029\\
	24	4.99951825885393\\
	24.25	4.99951825885393\\
	24.5	4.99951825885393\\
	24.75	4.99951825885393\\
	25	5\\
	25.25	5\\
	25.5	5\\
	25.75	5\\
	26	5\\
};
\addlegendentry{$H(A)+1$};

\addplot [color=mycolor3,dashed,line width=1.0pt]
table[row sep=crcr]{%
	5	1.62500552003331\\
	5.25	1.62500552003331\\
	5.5	1.62500552003331\\
	5.75	1.62500552003331\\
	6	1.62500552003331\\
	6.25	1.782876928828\\
	6.5	1.94853903271756\\
	6.75	2.08175846151053\\
	7	2.19198707891994\\
	7.25	2.28981034157059\\
	7.5	2.37627224944022\\
	7.75	2.45379194997537\\
	8	2.53299255339484\\
	8.25	2.60286634205178\\
	8.5	2.67721988838854\\
	8.75	2.75055116186289\\
	9	2.82218212020344\\
	9.25	2.89790288918891\\
	9.5	2.97121743349079\\
	9.75	3.04140175686331\\
	10	3.11526900065954\\
	10.25	3.18519909195096\\
	10.5	3.26712753683267\\
	10.75	3.33620244078796\\
	11	3.40874315485465\\
	11.25	3.48509805572241\\
	11.5	3.55534976079943\\
	11.75	3.63997584354267\\
	12	3.70677279868982\\
	12.25	3.78865545178417\\
	12.5	3.86252270507571\\
	12.75	3.94007739056627\\
	13	4.02160318099579\\
	13.25	4.1073836206079\\
	13.5	4.18231160259184\\
	13.75	4.26051648123379\\
	14	4.34210210125419\\
	14.25	4.42712314129943\\
	14.5	4.4976051718061\\
	14.75	4.57022370640973\\
	15	4.64487364021476\\
	15.25	4.72138517013066\\
	15.5	4.79950940034261\\
	15.75	4.85895817465942\\
	16	4.91893770241768\\
	16.25	4.97922343656377\\
	16.5	5.03954777467145\\
	16.75	5.09959652308648\\
	17	5.13929838235083\\
	17.25	5.19805387404306\\
	17.5	5.23649787140198\\
	17.75	5.2741912047721\\
	18	5.31097685428518\\
	18.25	5.34668776583955\\
	18.5	5.38114733728094\\
	18.75	5.41417010335418\\
	19	5.44556263452022\\
	19.25	5.4605852095352\\
	19.5	5.48915511105399\\
	19.75	5.50265043371992\\
	20	5.52793031311388\\
	20.25	5.53966191331352\\
	20.5	5.56117593100207\\
	20.75	5.57090559177726\\
	21	5.57991548896304\\
	21.25	5.58817981847353\\
	21.5	5.59567314707936\\
	21.75	5.60237049407994\\
	22	5.608247414588\\
	22.25	5.61328008404869\\
	22.5	5.61744538358718\\
	22.75	5.61744538358718\\
	23	5.62072098575262\\
	23.25	5.62072098575262\\
	23.5	5.6230854402029\\
	23.75	5.6230854402029\\
	24	5.62451825885393\\
	24.25	5.62451825885393\\
	24.5	5.62451825885393\\
	24.75	5.62451825885393\\
	25	5.625\\
	25.25	5.625\\
	25.5	5.625\\
	25.75	5.625\\
	26	5.625\\
};
\addlegendentry{$H(A)+1.625$};
\draw[->]  (axis cs:19,5.4487)--(axis cs:17,5.4487);
\node[] at (axis cs:16,5.4487) {$19$ dB};
\end{axis}
\end{tikzpicture}%
	\vspace{-2.7ex}
	\caption{AIRs ($\gmiHDD$) for $64$-QAM and transmission rate of the designed CM scheme ($H(A)+2\gamma$) with $2\gamma=1$ and $2\gamma=1.625$.} 
	\vspace{0.9ex}	
	\label{AIRfig} 
\end{figure}
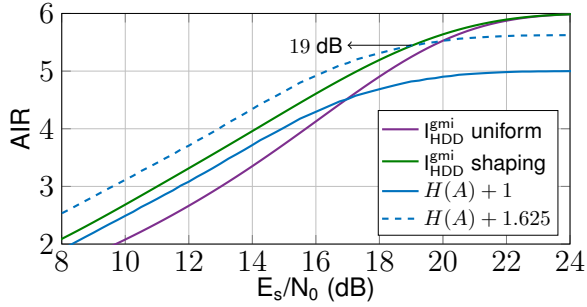
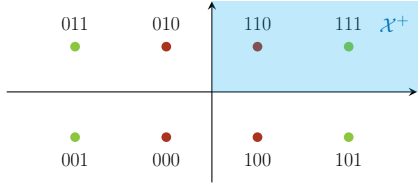
\begin{figure}[t]
	\begin{center}
		\begin{tikzpicture}[
		>=stealth,
		scale=.3,
		every node/.append style={transform shape}
		]		
		\tikzset{checknoderootw/.style={rectangle, draw, thick, minimum width=3mm, minimum height=3mm,fill=white}}	
		\tikzset{checknoderoot/.style={rectangle, draw, thick, minimum width=3mm, minimum height=3mm,fill=red!50!white}}
		\tikzset{checknode/.style={rectangle, draw, thick, minimum width=3mm, minimum height=3mm,fill=white}}
		\tikzset{root/.style={circle, draw, thick, minimum size=1mm, fill=cyan!60,opacity=0.8}}
		\tikzset{root1/.style={circle, draw, thick, minimum size=1mm, fill=blue!60,opacity=0.4}}
		\tikzset{root2/.style={circle, draw, thick, minimum size=1mm, fill=green!80,opacity=0.4}}
		\tikzset{root3/.style={circle, draw, thick, minimum size=1mm, fill=gray!80,opacity=0.8}}									
		\tikzset{bitnoderoot/.style={circle, draw=red, thick, minimum size=1mm, fill=red}}
		\tikzset{bitnode/.style={circle, draw, thick, minimum size=2mm, fill=white}}
		\tikzset{bitnode1/.style={circle, draw=Mahogany, thick, minimum size=2mm, fill=Mahogany}}	
		\tikzset{bitnode2/.style={circle, draw=LimeGreen, thick, minimum size=2mm, fill=LimeGreen}}			
		\tikzset{descr/.style={fill=white}}	
		\tikzset{Sourcebox2/.style={rectangle, thick, minimum width=9cm, minimum height=8cm}}
		\tikzset{Sourcebox1/.style={rectangle, thick, minimum width=9cm, minimum height=4cm}}
		\tikzset{descr/.style={fill=white}}
		
		\node[bitnode2] (x2) at (-6,-2) {};
		\node[bitnode2] (x3) at (-6,2) {};
		
		\node[bitnode1] (x6) at (-2,-2) {};
		\node[bitnode1] (x7) at (-2,2) {};
		
		\node[bitnode1] (x10) at (2,-2) {};
		\node[bitnode1] (x11) at (2,2) {};
		
		\node[bitnode2] (x14) at (6,-2) {};
		\node[bitnode2] (x15) at (6,2) {};
		
		\draw [->] (-9,0) -- (9,0);
		\draw [->] (0,-4) -- (0,4);   
		
		\node[] (l1) at (-6,3) {\Huge $011$};
		\node[] (l2) at (-6,-3) {\Huge $001$};
		\node[] (l3) at (-2,3) {\Huge $010$};	
		\node[] (l4) at (-2,-3) {\Huge $000$};
		\node[] (l5) at (2,3) {\Huge $110$};
		\node[] (l6) at (2,-3) {\Huge $100$};
		\node[] (l7) at (6,3) {\Huge $111$};	
		\node[] (l8) at (6,-3) {\Huge $101$};

		\node[Sourcebox1,fill=cyan, fill opacity=0.25] (top) at (4.53,2) {};		
		
		\node[] at (8,3) {\Huge \textcolor{RoyalBlue}{$\mathcal{X}^{+}$}};

		
		\end{tikzpicture}	
	\end{center}
	\vspace{-1.2ex}
	\caption{$8$-QAM constellation with BRGC labeling.}
	\label{fig8QAM} \vspace{0.5ex}
\end{figure}

\begin{figure*}[!t]
	\centering
	\vspace{-0.4cm}
	\scalebox{0.68}{
		\begin{tikzpicture}[>=latex']
		\tikzset{Source/.style={rectangle, draw, thick, minimum width=1.6cm, minimum height=1cm, rounded corners=2mm}}
		\tikzset{Sourcerealimag/.style={rectangle, draw, thick, minimum width=1cm, minimum height=0.3cm, rounded corners=2mm}}		
		\tikzset{Destination/.style={rectangle, draw, thick, minimum width=1.6cm, minimum height=1cm, rounded corners=2mm}}
		\tikzset{Noise/.style={circle, draw, thick, scale=0.5, minimum size=0.1mm}}
		\tikzset{descr/.style={fill=white}}
		\tikzset{Source1b/.style={rectangle, draw=black, thick, minimum width=0.5cm, minimum height=1.2cm, rounded corners=0.5mm}}
		\tikzset{Source2/.style={rectangle, draw, thick, minimum width=0.8cm, minimum height=1.2cm, rounded corners=0.5mm}}							
		\node[Source,fill=gray!20,thick] (C1) at (-22.2,3) {Uniform};
		\node[Source,fill=cyan!50!white] (C2) at (-19.5,3) {\Large $p^{\lambda^{*}}_{A}$};
		\node[] (B2) at (-19.7,,-5.8) {\large $A_1,...,A_n$};			
		\node[Source2,fill=red!30!white] (C3) at (-17.18,1.75) {\large $\Phi$};
		\node[Source,fill=cyan!50!white] (C4) at (-12.6,1.7) {Enc.};
		\node[Sourcerealimag,fill=red!30!white] (I) at (-12.6,3.3) {$j\mathcal{I(\cdot)}$};	
		\node[Sourcerealimag,fill=red!30!white] (R) at (-12.6,2.75) {$\mathcal{R(\cdot)}$};				
		\node[Source2,fill=red!30!white] (C5) at (-8.8,1.7) {\large $\Phi^{-1}$};
		\node[Noise] (C88) at (-6.7,3) {\huge $\times$};
		\node[Noise] (cc9) at (-2.8,3) {\huge $\times$};
		\node[Noise] (imag) at (-5.78,3.5) {\huge $\times$};
		\node[Noise] (imag1) at (-4.87,3) {\huge $+$};				
		\node[] (cc91) at (-5,3) {};
		\node[] (C7) at (-13.95,3) {};  
		\node[] (C77) at (-14.08,2.57) {};
		\node[] (C78) at (-14.08,3.45) {};				
		\node[] (B1) at (-20.1,2.1) {\large $\ua=(u_1,\ldots,u_{k-2\gamma n})$};
		\node[] (D1) at (-22.2,0.6) {};	
		\node[] (D2) at (-22.31,0.725) {};	
		\node[] (D3) at (-14.6,0.725) {};	
		
		\node[] (B1) at (-20.2,1.1) {\large $\us=(u_1,...,u_{2\gamma n})$};	
		\node[] (B2) at (-15.3,,-2.5) {\large $\b_1,...,\b_n$};	
		\node[] (B3) at (-11.5,,-2.5) {\large $p_1,...,p_{n(2-2\gamma)}$};	    	
		\node[] (B4) at (-8.75,,-3.2) {\large $s_1,...,s_{n}$};	
		\node[] (B4p) at (-7.6,,-1.4) {\large $s_{n+1},...,s_{2n}$};			
		\node[] (B5) at (-6.1,,-5.9) { $X_1,...,X_n$};		        	
		\node[] (C8) at (-18.2,3.12) {};	    
		\node[] (D4) at (-18.2,1.56) {};	
		\node[] (D5) at (-18.33,1.68) {};
		\node[] (D6) at (-17.45,1.68) {};

		\node[] (R1i) at (-12.3,3.5) {};
		\node[] (R1o) at (-5.95,3.5) {};

		\node[Source1b,fill=red!30!white] (M1) at (-15.5,1.7) {};
		\node[black] at (-15.5,1.78) {\small \vvv{\sffamily p/s}};	
		
		\draw [->] (-16.75,2.25) --  node [above] {\small $b_1$} (-15.75,2.25);
		\draw [->] (-16.75,2.25) --  node [below] {\small $\vdots$} (-15.75,2.25);		
		\draw [->] (-16.75,1.2) --  node [above] {\small $b_{m-2}$} (-15.75,1.2);
		
		\draw [->] (C1) -- (C2);
		\draw [-] (C2) -- (C7);		
		\draw [-] (C1) -- (D1);		
		\draw [-] (D2) -- (D3);	
		\draw [-] (C8) -- (D4);			
		\draw [->] (D5) -- (D6);	
		\draw [->] (M1) -- (C4);
		\draw [-] (C77) -- (C78);
		
		\draw [->] (R1i) -- (R1o);	
		\draw [-] (-12.12,2.8) -- (-8,2.8);		
		\draw [-] (-8,2.8) -- (-8,3);
		\draw [->] (-8,3) -- (-6.95,3);		
		
		\draw [->] (-14.07,2.7) -- (-13.1,2.7);
		\draw [->] (-14.07,3.32) -- (-13.1,3.32);							
		\draw [->] (-12.54,0.73) -- (-12.54,1.2);
		\draw [-] (-14.74,0.73) -- (-8.84,0.73);
		\draw [->] (-8.85,0.73) -- (-8.85,1.1);				
		\draw [->] (C4) -- (C5);	
		\draw [-] (-8.35,2.1) -- (-6.7,2.1);
		\draw [->] (-6.7,2.1) -- (-6.7,2.7);	
		\draw [-] (-8.35,1.34) -- (-5.8,1.34);
		\draw [->] (-5.8,1.34) -- (-5.8,3.24);		
		\draw [->] (-6.7,2.1) -- (-6.7,2.7);			
		\draw [->] (C88) -- (cc91);	
		\draw [->] (imag1) -- (cc9);	
		\draw [-] (imag) -- (-4.85,3.5);	
		\draw [->] (-4.85,3.5) -- (-4.85,3.25);					
		\draw [->] (-2.8,1.7) -- (-2.8,2.75);	
		\node[] (D10) at (-2.75,1.5) {$\Lambda^{*}$};
		\node[cloud, cloud puffs=12.7, cloud ignores aspect, minimum width=0.2cm, minimum height=0.2cm, align=center, draw,fill=gray!20,thick] (cloud) at (-1.3,3) {FOC};		
		\draw [->] (cc9) -- (cloud);																							
		
		\end{tikzpicture}
	}
	
	\caption{Block diagram of the proposed CM scheme. $\mathcal{R(\cdot)}$ takes the real part while $\mathcal{I(\cdot)}$ takes the imaginary part.}
	\label{figsystemmodel}
	\vspace{-0.4ex}
\end{figure*}

\begin{table}[!t]
	
	\renewcommand{\tabcolsep}{0.15cm}
	\caption {Parameters of the designed staircase codes for $v=10$, $t=3$, and $64$-QAM modulation} \label{table:feasibleparam}
	\vspace{-0.25cm}
	\begin{center}
		\vspace{-1ex}
		\begin{center}\scalebox{0.8}{\begin{tabular}{ccccccc}
				\arrayrulecolor{black}\hline
				\toprule

				$2\gamma$ & $1.625$ & $1.5313$ & $1.4444$ & $1.3023$ & $1.1429$  & $1$\\
				\midrule
				$s$ & $63$ & $255$ & $375$ & $507$ & $603$ & $663$ \\
				\midrule				
				$\Rs$ & $0.9375$ & $0.9219$ & $0.9074$ & $0.8837$ & $0.8571$ & $0.8333$ \\		
				
				\hline
				\toprule                			
			\end{tabular} }\end{center}
		\end{center}
		\vspace{-0.8cm}
	\end{table}
\section{\AGiA{Proposed Coded Modulation Scheme}}
\AGiA{The proposed CM scheme is shown in Fig.~\ref{figsystemmodel}}. Let $\u=(u_1,...,u_{k})$ be the information block, of length $k$ bits, generated by a uniform source. $\u$ is split into two vectors $\us$ and $\ua$, of lengths $2\gamma n$ and $k-2\gamma n$, respectively. $\ua$ is used as the input of the shaping block (also called distribution matcher), shown with blue color in Fig.~\ref{figsystemmodel}, where the output is from the set $\mathcal{X}^{+}$ with distribution $p^{\lambda^{*}}_{A}$. Here, we consider the constant composition distribution matcher\cite{Schulte_2016}. 
\AGiA{The ${A_i}$'s are transformed to length $m-2$-bit vectors $\b_i$ by the mapper $\Phi$, which uses the BRGC labeling}, and the resulting bits are concatenated using a parallel to serial converter. Therefore, $n(m-2)$ bits are contained in $\b_1,...,\b_n$. 

\AGiA{We use staircase codes with binary systematic BCH codes as component codes. Let $\mathcal{C}$ be an $(\nc,\kc)$ shortened BCH code of (even) code length $\nc=2^v-1-s$ and dimension $\kc=2^v-vt-1-s$ constructed over the Galois field $\text{GF}(2^v)$, where $s$ is the shortening length and $t$ the error correcting capability of the code}. A shortened BCH code is thus completely specified by the triple $(v,t,s)$. A staircase code with $(\nc,\kc)$ component codes is defined as the set of all $\frac{\nc}{2} \times \frac{\nc}{2}$ matrices $\Bmat_\text{i}$, $\text{i}=1,2,\ldots$, such that each row of the matrix $[\Bmat_\text{i-1}^\rmT,\Bmat_\text{i}]$ is a valid codeword of $\mathcal{C}$\cite{staircase_frank}. 

\AGiA{We set the rate of the staircase code to 
\begin{align}
\label{eq:rate}
\Rs = 1 - \frac{2(\nc-\kc)}{\nc} =  \frac{{m - 2 + 2\gamma }}{m},
\end{align}}%
where $1/2 \le \gamma  < 1$ is a tuning parameter that can be used to select the rate of the staircase code and subsequently the spectral efficiency (SE) of the CM scheme. The sequences $\b_1,...,\b_n$ and $\us$ form the information bits of one staircase block, with $n=\frac{(\kc-\nc/2)\cdot(\nc/2)}{m-2+2\gamma}$. This specific value for $n$ is due to \AGiA{the rate $\Rs$ in \eqref{eq:rate}} and the structure of the staircase code where each component code is spread over two consecutive blocks. \AGiA{We select $(\nu,t,s)$ such that the number of parities in each} staircase block is $n(2-2\gamma)$, which should be an integer number. Table ~\ref{table:feasibleparam} summarizes some of the designed code parameters for $\nu=10$ and $t=3$. One can easily show that the parity bits have a \GL{distribution that closely approximates the uniform one} \cite{georg_tcom}. The uniform data bits $\us$ are attached to the parity bits $p_1,...,p_{n(2-2\gamma)}$ and transformed to signs \GL{($0 \mapsto -1$ and $1 \mapsto +1$) which are finally multiplied by the real and imaginary parts of $A_1,...,A_n$.} The $X_1,...,X_n$ are then rescaled by $\Lambda^{*}$ to maintain the power constraint \GL{before} being sent \GL{over} the FOC. At the receiver, decoding of the staircase code is performed using bounded distance decoding and extrinsic message passing\cite{jian2013iterative}. 

One can easily show that \AGiA{the transmission rate} of the proposed CM scheme is $H(A)+2\gamma$. Since vanishing error probability is only possible if $H(A)+2\gamma<\gmiHDD$, by crossing the curves $H(A)+2\gamma$ and $\gmiHDD$, one can find the feasible SNR region where the CM scheme can operate. In Fig.~\ref{AIRfig}, for $2\gamma=1$ all SNRs to the right of the curve $H(A)+1$ are feasible while for $2\gamma=1.625$ only the SNRs larger than $19$ dB are feasible. 
\begin{figure}[t] \centering 
	\setlength\figureheight{4.3cm}
	\setlength\figurewidth{8.8cm}  
%
%
\definecolor{mycolor1}{rgb}{0.49412,0.18431,0.55686}%
\definecolor{mycolor2}{rgb}{0.00000,0.49804,0.00000}%
\definecolor{mycolor3}{rgb}{0.00000,0.44706,0.74118}%
\definecolor{mycolor4}{rgb}{0.49412,0.18431,0.55686}%
\begin{tikzpicture}[scale=.8]
\pgfplotsset{every tick label/.append style={font=\Large}}
\begin{axis}[%
width=0.95\figurewidth,
height=1.47\figureheight,
at={(0\figurewidth,0\figureheight)},
scale only axis,
xmin=16,
xmax=21.5,
ymin=4.3,
ymax=5.8,
ylabel={\large spectral eff. (bits/symbol)},
xlabel={\large {$\text{E}_\text{s}\text{/N}_\text{0}$ (dB)}},
ylabel style={yshift=-0.25cm},
xlabel style={yshift=0.1cm},
xmajorgrids,
ymajorgrids,
axis background/.style={fill=white},
legend style={at={(1.002,-0.003)},anchor=south east,row sep=0pt, legend cell align=left,align=left,draw=black, fill=white, inner sep=1.3pt, outer sep=0.9pt}
]
\addplot [color=mycolor4,solid,line width=1.0pt]
table[row sep=crcr]{%
	5	0.999062623942368\\
	5.25	1.04161939853981\\
	5.5	1.08518319110512\\
	5.75	1.12982015705296\\
	6	1.17561171097448\\
	6.25	1.22251454568213\\
	6.5	1.27055866902754\\
	6.75	1.31977116414262\\
	7	1.37017527873563\\
	7.25	1.42196573422968\\
	7.5	1.474998580681\\
	7.75	1.52927589274995\\
	8	1.5848900270839\\
	8.25	1.64190047371988\\
	8.5	1.70016662725648\\
	8.75	1.75984076197366\\
	9	1.82087158532825\\
	9.25	1.8832208290044\\
	9.5	1.94702539426181\\
	9.75	2.01213725898699\\
	10	2.07872181166046\\
	10.25	2.14668296304102\\
	10.5	2.21605960502699\\
	10.75	2.2869616229617\\
	11	2.35928780160918\\
	11.25	2.43308955979793\\
	11.5	2.5085178111805\\
	11.75	2.58548750223494\\
	12	2.66403759299672\\
	12.25	2.7442020608791\\
	12.5	2.82600565680176\\
	12.75	2.90946161261522\\
	13	2.99456925754282\\
	13.25	3.08131162659362\\
	13.5	3.169762487932\\
	13.75	3.25978698794393\\
	14	3.35134882577708\\
	14.25	3.44442914743305\\
	14.5	3.53890215696059\\
	14.75	3.63464264476682\\
	15	3.73154062027921\\
	15.25	3.82944839339683\\
	15.5	3.92818129231541\\
	15.75	4.02751907716647\\
	16	4.12720803964009\\
	16.25	4.22705660593868\\
	16.5	4.32673101064465\\
	16.75	4.42596985615945\\
	17	4.52447405460975\\
	17.25	4.62191301301898\\
	17.5	4.7179314431442\\
	17.75	4.81220225806347\\
	18	4.90436921942489\\
	18.25	4.99409855943148\\
	18.5	5.08104338335846\\
	18.75	5.16486778164169\\
	19	5.24526587284423\\
	19.25	5.32195520307841\\
	19.5	5.39466520230525\\
	19.75	5.46316166682473\\
	20	5.52726242116612\\
	20.25	5.58681822952208\\
	20.5	5.64173697706689\\
	20.75	5.69196673426747\\
	21	5.73752042354205\\
	21.25	5.77845380319505\\
	21.5	5.81488182575306\\
	21.75	5.84696918000668\\
	22	5.87492409186845\\
	22.25	5.89899913389514\\
	22.5	5.9194774580799\\
	22.75	5.93667048961884\\
	23	5.95090514777848\\
	23.25	5.96251796794189\\
	23.5	5.97184450198895\\
	23.75	5.97921155913974\\
	24	5.98492876943471\\
	24.25	5.98928331387043\\
	24.5	5.99253473417472\\
	24.75	5.99491183370014\\
	25	5.99661129991762\\
};
\addlegendentry{$\gmiHDD$ uniform};

\addplot [color=mycolor2,solid,line width=1.0pt]
table[row sep=crcr]{%
	5	1.46579892317654\\
	5.25	1.51178911680456\\
	5.5	1.55619123984638\\
	5.75	1.59882968188983\\
	6	1.63953427325279\\
	6.25	1.68000853670076\\
	6.5	1.72577949084277\\
	6.75	1.77689532479202\\
	7	1.83272060466621\\
	7.25	1.89255005724396\\
	7.5	1.95571822430114\\
	7.75	2.02166184994724\\
	8	2.08990018738181\\
	8.25	2.1600553890047\\
	8.5	2.2318070653719\\
	8.75	2.304916488661\\
	9	2.37916205387676\\
	9.25	2.45438436936732\\
	9.5	2.5304187433945\\
	9.75	2.60715269627337\\
	10	2.68446550131693\\
	10.25	2.76226187311712\\
	10.5	2.84045015770386\\
	10.75	2.91896869850708\\
	11	2.99773143907823\\
	11.25	3.07671422861147\\
	11.5	3.15586817838582\\
	11.75	3.23517892321605\\
	12	3.31467672592584\\
	12.25	3.39435455116865\\
	12.5	3.47423829795137\\
	12.75	3.5543838786503\\
	13	3.63481492304937\\
	13.25	3.71555889895922\\
	13.5	3.79661804137871\\
	13.75	3.87795188105578\\
	14	3.95955103232285\\
	14.25	4.04130553418989\\
	14.5	4.12310827933874\\
	14.75	4.20485845514854\\
	15	4.28642427445235\\
	15.25	4.36767479289973\\
	15.5	4.44840442014369\\
	15.75	4.52852864511378\\
	16	4.60782481644679\\
	16.25	4.68615062673721\\
	16.5	4.76333932398564\\
	16.75	4.83919931168793\\
	17	4.91364046986271\\
	17.25	4.98639190833142\\
	17.5	5.05742527251708\\
	17.75	5.12647157663465\\
	18	5.19340459798565\\
	18.25	5.2581069764854\\
	18.5	5.3204322365192\\
	18.75	5.38023999842817\\
	19	5.43737553927009\\
	19.25	5.49181532197591\\
	19.5	5.54345964303734\\
	19.75	5.59206986965365\\
	20	5.63785184467143\\
	20.25	5.68050461548196\\
	20.5	5.72005698814792\\
	20.75	5.75666720425817\\
	21	5.7901578693714\\
	21.25	5.82058855324342\\
	21.5	5.84803707424999\\
	21.75	5.87258686484818\\
	22	5.8943312455479\\
	22.25	5.91337116309528\\
	22.5	5.92981285691288\\
	22.75	5.94397668986422\\
	23	5.95595535683365\\
	23.25	5.96576921435593\\
	23.5	5.97402262310257\\
	23.75	5.98049623878471\\
	24	5.98571329436761\\
	24.25	5.98975673418029\\
	24.5	5.99275855496915\\
	24.75	5.99493988371216\\
	25	5.99661129991762\\
	25.25	5.99779782206388\\
	25.5	5.99860562755232\\
	25.75	5.99914113721467\\
	26	5.9994862299573\\
};
\addlegendentry{$\gmiHDD$ shaping};

\addplot [color=mycolor2,line width=2.0pt,mark size=4.0pt,only marks,mark=x,mark options={solid}]
table[row sep=crcr]{%
	19.832	5.5106\\
	19.32	5.3737\\
	18.858	5.2394\\
	18.1	4.9917\\
	17.3	4.6955\\
	16.5	4.38\\
};
\addlegendentry{Sim. $p_X^{{\lambda ^*}}$ dist.};

\addplot [color=mycolor1,line width=2.0pt,mark size=4.0pt,only marks,mark=x,mark options={solid}]
table[row sep=crcr]{%
	20.91	5.625\\
	20.61	5.5313\\
	20.36	5.4444\\
	20	5.3023\\
	19.68	5.1429\\
	19.42	5\\
};
\addlegendentry{Sim. uniform dist.};

\addplot [color=mycolor3,dashed,line width=1.0pt]
table[row sep=crcr]{%
	5	1.53130552003331\\
	5.25	1.53130552003331\\
	5.5	1.53130552003331\\
	5.75	1.53130552003331\\
	6	1.53130552003331\\
	6.25	1.689176928828\\
	6.5	1.85483903271756\\
	6.75	1.98805846151053\\
	7	2.09828707891994\\
	7.25	2.19611034157059\\
	7.5	2.28257224944022\\
	7.75	2.36009194997537\\
	8	2.43929255339484\\
	8.25	2.50916634205178\\
	8.5	2.58351988838854\\
	8.75	2.65685116186289\\
	9	2.72848212020344\\
	9.25	2.80420288918891\\
	9.5	2.87751743349079\\
	9.75	2.94770175686331\\
	10	3.02156900065954\\
	10.25	3.09149909195096\\
	10.5	3.17342753683267\\
	10.75	3.24250244078796\\
	11	3.31504315485465\\
	11.25	3.39139805572241\\
	11.5	3.46164976079943\\
	11.75	3.54627584354267\\
	12	3.61307279868982\\
	12.25	3.69495545178417\\
	12.5	3.76882270507571\\
	12.75	3.84637739056627\\
	13	3.92790318099579\\
	13.25	4.0136836206079\\
	13.5	4.08861160259184\\
	13.75	4.16681648123379\\
	14	4.24840210125419\\
	14.25	4.33342314129943\\
	14.5	4.4039051718061\\
	14.75	4.47652370640973\\
	15	4.55117364021476\\
	15.25	4.62768517013066\\
	15.5	4.70580940034261\\
	15.75	4.76525817465942\\
	16	4.82523770241768\\
	16.25	4.88552343656377\\
	16.5	4.94584777467145\\
	16.75	5.00589652308648\\
	17	5.04559838235083\\
	17.25	5.10435387404306\\
	17.5	5.14279787140198\\
	17.75	5.1804912047721\\
	18	5.21727685428518\\
	18.25	5.25298776583955\\
	18.5	5.28744733728094\\
	18.75	5.32047010335418\\
	19	5.35186263452022\\
	19.25	5.3668852095352\\
	19.5	5.39545511105399\\
	19.75	5.40895043371992\\
	20	5.43423031311388\\
	20.25	5.44596191331352\\
	20.5	5.46747593100207\\
	20.75	5.47720559177726\\
	21	5.48621548896304\\
	21.25	5.49447981847353\\
	21.5	5.50197314707936\\
	21.75	5.50867049407994\\
	22	5.514547414588\\
	22.25	5.51958008404869\\
	22.5	5.52374538358718\\
	22.75	5.52374538358718\\
	23	5.52702098575262\\
	23.25	5.52702098575262\\
	23.5	5.5293854402029\\
	23.75	5.5293854402029\\
	24	5.53081825885393\\
	24.25	5.53081825885393\\
	24.5	5.53081825885393\\
	24.75	5.53081825885393\\
	25	5.5313\\
	25.25	5.5313\\
	25.5	5.5313\\
	25.75	5.5313\\
	26	5.5313\\
};
\addlegendentry{$H(A)+1.5313$};

\addplot [color=mycolor2,line width=2.0pt,mark size=4.0pt,only marks,mark=x,mark options={solid},opacity=0.45,forget plot]
table[row sep=crcr]{%
    19.55	 5.3879\\
	19.75   5.4014\\
	20	    5.4265\\
};

\draw[<->,thick]  (axis cs:18.08,4.95)--(axis cs:19.42,4.95);
\node[] at (axis cs:18.7,4.91) {$1.32$ dB};

\end{axis}
\end{tikzpicture}%
	\vspace{-2ex}
	\caption{AIRs ($\gmiHDD$) and simulation results for the proposed CM scheme with $64$-QAM.} 
	\vspace{-1.2ex}	
	\label{AIRfig1} 
\end{figure}
\section{Simulation Results}
We simulated the performance of the proposed CM scheme using the designed codes summarized in Table~\ref{table:feasibleparam}. 
We set the target block error probability to $\text{P}_\text{e}=3\times10^{-3}$, corresponding to a bit error probability of roughly $10^{-7}$, and by means of simulations \AGiA{we find the minimum required SNR such that} the target $\text{P}_\text{e}$ is met. This is shown with dark green crosses in Fig.~\ref{AIRfig1}. We remark that each point corresponds to a specific $\gamma$ and it is on the curve $H(A)+2\gamma$. This means that the points on the curve $H(A)+2\gamma$ corresponding to larger SNRs can also meet the target performance. For example, for $2\gamma=1.5313$, some other possible operational points are shown with faint green crosses.
However, since by increasing the SNR $H(A)+2\gamma$ becomes flat, at some point one needs to change  $\gamma$ (change the component code) to be as close as possible to the AIR $\gmiHDD$. We remark that one can vary the \AGiA{SE} from $4.38$ to $5.5$ bits/symbol using a single code and decoder by simply changing the shaping distribution and shortening of the component code. \AGiA{Remarkably, our CM scheme with (finite-length) staircase codes achieves performances that are better than the AIRs (which assume ideal codes) of the CM scheme with uniform distribution}. For the sake of comparison, we also simulated the performance of the system using the same codes with uniform distribution, i.e.  (purple crosses). \AGiA{The proposed probabilistically-shaped CM scheme} achieves significantly better performance compared to the uniform scheme. For a SE of $5$ bits/symbol, the gain is $1.32$ dB. 
\vspace{-0.1cm}
\section{Conclusion}
\AGiA{We proposed a CM scheme based on binary staircase codes with HDD for high-speed fiber-optic communications. The proposed scheme uses probabilistic shaping to maximize the AIR of the system. Remarkably, the performance of the proposed CM scheme with  staircase codes is better than the AIR (i.e., assuming ideal codes) of the CM scheme with uniform distribution. Gains over $1$ dB are observed compared to the performance of the baseline CM scheme with staircase codes and uniform distribution. Furthermore, the proposed CM scheme achieves performance within \AS{$0.5$--$1.2$} dB of the AIR for a wide range of SEs using only a single staircase code, which greatly reduces the decoder complexity.}  

\vspace{-0.2cm}
\section{Acknowledgment}
\footnotesize{This work was financially supported by the Knut and Alice Wallenberg Foundation.}

\vspace{-0.2cm}


\end{document}